\documentclass[a4paper,fleqn,usenatbib]{mnras}

\usepackage[T1]{fontenc}
\usepackage{ae,aecompl}

\usepackage{graphicx}	
\usepackage{amsmath}	
\usepackage{amssymb}	

\usepackage[utf8]{inputenc}
\usepackage{hyperref}
\usepackage{microtype}
\usepackage{natbib}
\usepackage{txfonts}
\usepackage{color}
\usepackage{multirow}
\usepackage{xspace}
\usepackage{hyperref}
\bibpunct{(}{)}{;}{a}{}{,} 
\newcommand{\src}{{Swift~J0243.6$+$6124}\xspace}

\def\flux{\,erg s$^{-1}$ cm$^{-2}$\xspace}

\def\lum{\,erg\,s$^{-1}$\xspace}

\def\gro{GRO~J1744$-$28}

\newcommand{\subt}[1]{_{\text{#1}}}

\newcommand {\be}{\begin {equation}}
\newcommand {\ee}{\end {equation}}
\newcommand {\beq}{\begin {eqnarray}}
\newcommand {\eeq}{\end {eqnarray}}

\newcommand{\powten}[1]{\times 10^{#1}}

\title{Hot disk of the \src revealed by Insight-HXMT}
\author[V.\,Doroshenko et al]{
V.\,Doroshenko$^{2,5,*}$,
S.N. Zhang$^{1,3,*}$,
A.\,Santangelo$^{2}$,
L.~\,Ji$^{2}$,
S.\,Tsygankov$^{4,5}$,
\newauthor
A.\,Mushtukov$^{6,7,5}$,
L.J.\,Qu$^{1}$,
S.\,Zhang$^{1}$,
M.Y.\,Ge$^{1}$
% ...
\\
\\
$^1$Key Laboratory for Particle Astrophysics, Institute of High Energy Physics, Chinese Academy of Sciences,\\ 19B Yuquan Road, Beijing 100049, People’s Republic of China\\
$^2$Institut für Astronomie und Astrophysik, Sand 1, 72076 Tübingen, Germany\\
$^3$University of Chinese Academy of Sciences, Chinese Academy of Sciences, Beijing 100049, People’s Republic of China\\
$^4$Department of Physics and Astronomy, FI-20014 University of Turku, Turku, Finland\\
$^5$Space Research Institute of the Russian Academy of Sciences, Profsoyuznaya Str. 84/32, Moscow 117997, Russia\\
$^6$Leiden Observatory, Leiden University, NL-2300RA Leiden, The Netherlands\\
$^7$Anton Pannekoek Institute, University of Amsterdam, Science Park 904, NL-1098 XH Amsterdam, the Netherlands\\
$^8$Department of Astronomy, Beijing Normal University, Beijing 100088, People’s Republic of China\\
$^9$Department of Physics, Tsinghua University, Beijing 100084, People’s Republic of China}
\begin{document}
\maketitle
\begin{abstract}
We report on analysis of observations of the bright transient X-ray pulsar \src obtained during its 2017-2018 giant outburst with Insight-HXMT, \emph{NuSTAR}, and \textit{Swift} observatories. We focus on the discovery of a sharp state transition of the timing and spectral properties of the source at super-Eddington accretion rates, which we associate with the transition of the accretion disk to a radiation pressure dominated (RPD) state, the first ever directly observed for magnetized neutron star. This transition occurs at slightly higher luminosity compared to already reported transition of the source from sub- to super-critical accretion regime associate with onset of an accretion column. We argue that this scenario can only be realized for comparatively weakly magnetized neutron star, not dissimilar to other ultra-luminous X-ray pulsars (ULPs), which accrete at similar rates. Further evidence for this conclusion is provided by the non-detection of the transition to the propeller state in quiescence which strongly implies compact magnetosphere and thus rules out magnetar-like fields.
\\
\\
\end{abstract}
\section{Introduction}
The transient binary X-ray pulsar (XRP) \src was first discovered on Oct.~3,
2017 \citep{Kennea17}, and for a few months exhibited one of the brightest
outbursts ever observed from a transient XRP with a peak flux of
$\sim3\times10^{-7}$\flux. Pulsations with steadily decreasing period of about
9.8\,s \citep{Kennea17,Jenke17} implied accretion onto the neutron star from
its Be companion \citep{opt_count}. Despite the identified counterpart, the
distance to the source remains uncertain. For the rest of the paper we adopt a
distance of 6.8\,kpc as reported in \cite{Bailer18} based on Gaia~DR2 parallax
measurements of the companion star. We emphasize, however, that even the lower
limit of $\sim5.5$\,kpc obtained from Gaia's measurements and based on the
observed spin-up rate of the neutron star \citep{Eijnden18,Doroshenko18},
implies that the source is an ultraluminous pulsar with X-ray luminosity larger
than $\sim10^{39}$\,erg\,s$^{-1}$ \citep{Tsygankov18}.

Another interesting feature of the source was unveiled by radio observations,
which confirmed emission correlated with the X-ray flux close to the
peak of the outburst. This was associated by \cite{Eijnden18} with an evolving
jet, one of the first ever observed from magnetized neutron stars, although
radio emission was also observed at later stages of the outburst at
significantly lower luminosities \citep{Eijnden18_her,Eijnden18}.

Similarly to other ULPs the magnetic field of \src is not measured directly. No
evidence for a cyclotron resonant scattering feature (CRSF, see
\cite{2019A&A...622A..61S} for a recent review), which would allow to
unambiguously measure the magnetic field of the neutron star, has been reported
in the 3-80\,keV energy band observed by \textit{NuSTAR}
\citep{2018MNRAS.474.4432J,Tao19}, nor in the Insight-HXMT observations in the
2-150\,keV energy range \citep{Zhang19}. Other arguments thus had to be invoked
to estimate the source's magnetic field. The observed change of the
pulse-profile shape in the soft band, and of the pulsed fraction in the hard
band prompted \cite{Doroshenko18} to conclude that the source switched from the
sub- to the super- critical accretion regime \citep{Basko76} at a luminosity of
$\sim10^{38}$\,erg\,s$^{-1}$, implying that the neutron star has a magnetic
field of $\sim10^{13}$\,G, i.e., slightly higher than that usual for accreting
pulsars. A similar conclusion was reached by \cite{nicer18} based on the
monitoring of the source with NICER and \textit{Fermi}/GBM, which revealed
peculiar features in the dependence of source's X-ray colours on luminosity
both in the soft and hard band. Both features, despite the significant
difference of the luminosity at which they occured, were associated with the
transition from the sub- to the super- critical accretion and with onset of an
accretion column \citep{nicer18}. Independently, since the source continued to
accrete at a luminosity of $\sim6\times10^{35}$\lum without switching to the
``propeller'' regime \citep{Illarionov75}, \cite{Tsygankov18} estimated an
upper limit for the magnetic field of $\sim6\times10^{12}$\,G, a value barely
consistent with other estimates.

Here we present the timing analysis of observations of \src obtained during the
source's 2017-2018 outburst with the Insight-HXMT satellite
\citep{2007NuPhS.166..131L}. We also present observations taken during the
following quiescence phase with the \emph{NuSTAR} mission
\citep{2013ApJ...770..103H}. Based on this analysis, we report the detection of
a striking change of the aperiodic variability properties, pulse profile
morphology, and energy spectrum of the source at a luminosity of
$L_x\sim4.4\times10^{38}$\lum, which we associate with the transition of the
inner regions of the accretion disk from the standard gas pressure dominated
(GPD) to the radiation pressure dominated (RPD) state. At this point the disc
pushes close enough to the neutron star to make local luminosity of inner disc
regions large enough to dynamically affect disc structure by radiative pressure
\citep{shakura73}. This affects the observed X-ray spectrum, aperiodic
variability originating within the disc, and coupling of the disc with
magnetosphere of the neutron star reflected by change of the observed pulse
profiles from the pulsar. In addition, also the already reported transition
from sub- to super-critical accretion regime is observed.

A second important and rather surprising result of our analysis is the
discovery of accretion powered X-ray emission from the source in deep
quiescence, months after the main outburst, at a luminosity as low as
$\sim3\times10^{34}$\lum. A coherent explanation of our findings and of the
phenomenology already reported in literature unambiguously shows that the
accretion disk extends unusually close to the compact object both in quiescence
and outburst. We therefore conclude that the source's magnetic field is likely
weaker than previously argued. The rest of the paper is organized as follows:
the details of the analysis are presented in section \ref{sec:obs} which is
followed by interpretation of individual observational findings and their
implications in section \ref{sec:disc}, and conclusions in section
\ref{sec:sum}.

\section{Observations and data analysis}
\label{sec:obs}
\subsection{\textit{NuSTAR} and \textit{Swift}/XRT}
The source had been observed with \textit{NuSTAR} on several occasions
\citep{Tao19}, however, here we focus exclusively on the most recent
observation after transition to quiescence. The observation was specifically
aimed to improve the upper limit of the magnetic field by \cite{Tsygankov18},
and was conducted following the rapid decline of the flux revealed by
\textit{Swift}/BAT monitoring of the source on MJD~58557 (80\,ks exposure,
observation id. 90501310002).

The data reduction was carried out using standard procedures using the
\emph{nustardas\_06Jul17\_v1.8.0} package and most updated calibration files.
Since the source has been clearly detected in images of both \textit{NuSTAR}
telescopes, to improve counting statistics we combined data of both units to
perform timing and spectral analysis. In particular, we concatenated
filtered event lists for timing, while spectra from the two units were
extracted and modeled independently and co-added using the \texttt{addspec}
tool for plotting only.

To extract source and background spectra and lightcurves, we used circular
extraction regions with radii of 80$^{\prime\prime}$ and 200$^{\prime\prime}$
centered on the source and close to the edge of the field of view on the same
detector chip. The source extraction radius was optimized to achieve the best
signal-to noise ratio above 40\,keV using the procedure described in
\cite{Vybornov18}. The lightcurves were also corrected to solar barycenter and
for the motion of the binary system using the ephemerides provided by
Fermi~GBM\footnote{\url{https://gammaray.nsstc.nasa.gov/gbm/science/pulsars/ligh
tcurves/swiftj0243.html}}.

\begin{figure}
    \centering
        \includegraphics[width=\columnwidth]{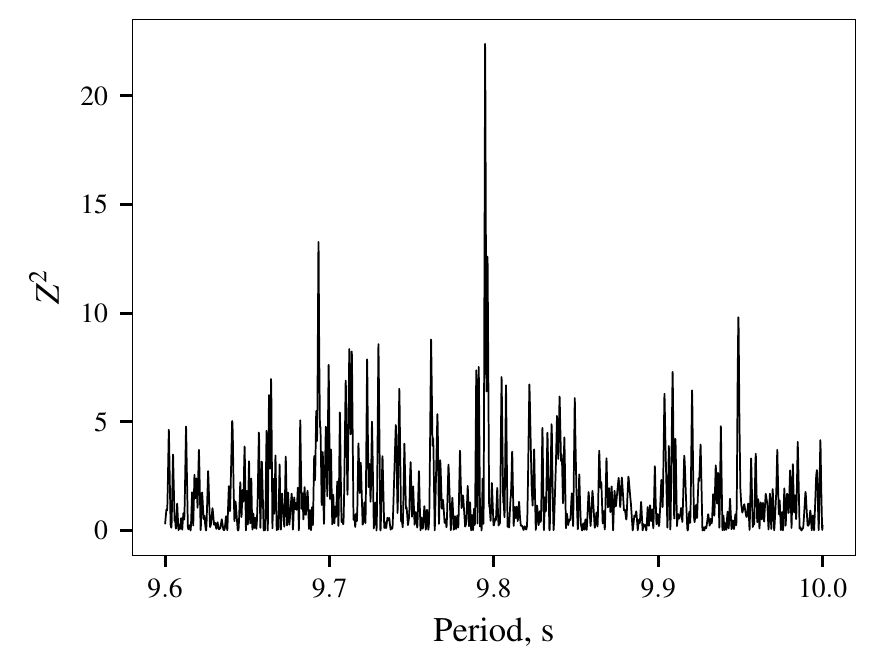}
        \includegraphics[width=\columnwidth]{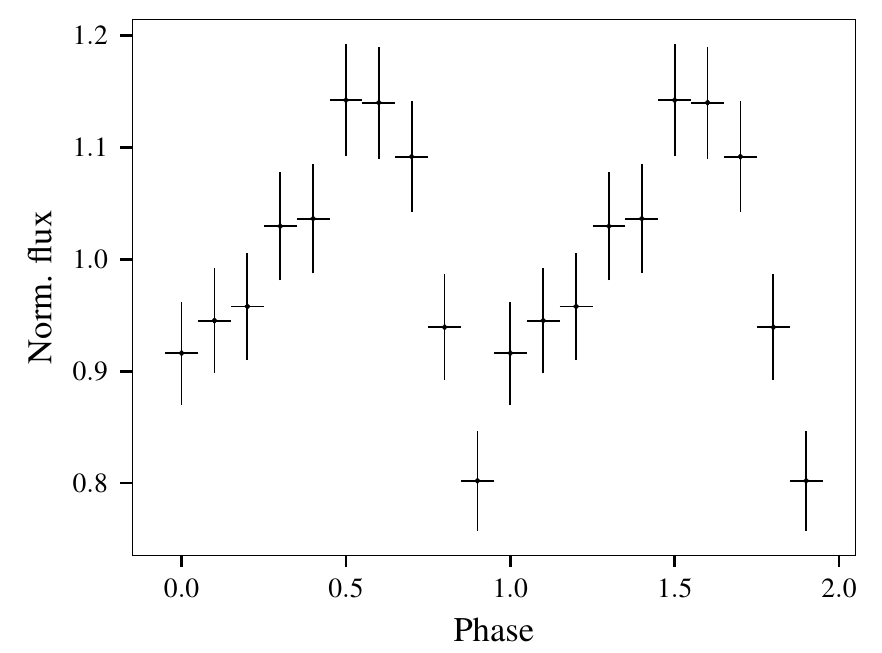}
    \caption{\textit{Top:} A periodogram for event arrvial times for \textit{NuSTAR} observation 90501310002 (80$^{\prime\prime}$ extraction radius, 3-80\,keV band). \textit{Bottom:} Background-subtracted lightcurve for \textit{NuSTAR} observation 90501310002 in 3-80\,keV band folded with best fit period.
    The pulse profile is plotted two times for clarity.}
    \label{fig:zstat}
\end{figure}

The main goal of the observation was to determine whether accretion continues
and the source continues to pulsate. To search for pulsations we used $Z^2$
statistics \citep{deJager89} on event data and the procedure described in
\cite{Doroshenko15} to avoid loss of sensitivity due to binning of the
lightcurves. A significant peak with $Z^2_1\sim22.4$ around the expected
frequency ($P\sim9.7946$\,s) was detected as shown in Fig.~\ref{fig:zstat}.
Even without considering that the most significant peak appears exactly at the
expected spin frequency, the corresponding chance detection probability for one
harmonic and $10^{4}$ trial frequencies is $\sim10^{-11}$ \citep{deJager89},
i.e., the peak is highly significant. The folded background-subtracted
3-80\,keV lightcurve reveals a single peaked pulse profile with pulsed fraction
of $\sim20$\% which strongly suggests that the source continues to accrete (see
Fig.~\ref{fig:zstat}).

\begin{figure}
    \centering
        \includegraphics[width=\columnwidth]{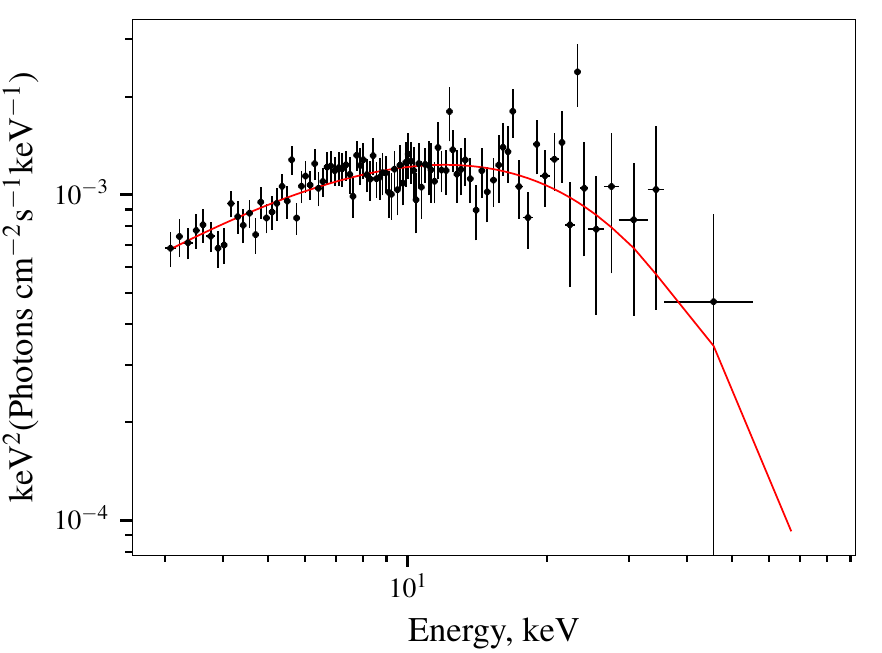}
    \caption{Broadband energy spectrum of the source as observed by \textit{NuSTAR} in quiescence unfolded with best-fit model and multiplied by energy squared.}
    \label{fig:qspe}
\end{figure}

The energy spectrum of the source in quiescence remains hard and is well
described with a cutoff power-law (\texttt{cutoffpl} in Xspec) with no
additional components, which also points to continued accretion (see
Fig.~\ref{fig:qspe}). Interstellar absorption was not required by the
fit but was included in the model fixed to interstellar value of
$9\times10^{21}$\,atoms\,cm$^{-2}$ for consistency with \textit{Swift/XRT}.
The best-fit statistics of $\chi^2\sim99$ for 104 degrees of freedom indicates
the good quality of the fit. Best fit photon index and cutoff energy are 1.0(1)
and 12.5(2)\,keV respectively ($1\sigma$ confidence level). The bolometric
model flux turns out to be $5.5(3)\times10^{-12}$\flux, a factor of twenty
smaller compared to previously published limit \citep{Tsygankov18}.

\begin{figure}
    \centering
        \includegraphics[width=\columnwidth]{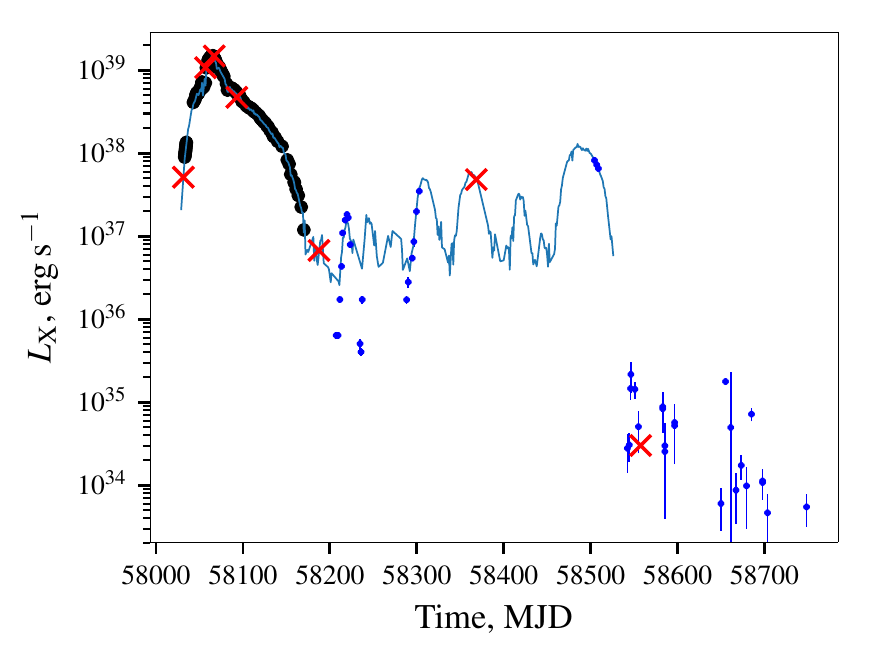}
    \caption{Long-term bolometric lightcurve of \src as observed by Insight-HXMT (black points), \textit{Swift}/BAT (blue line), \textit{Swift}/XRT (blue points), and \textit{NuSTAR} (red crosses). 
    The dimmest \textit{NuSTAR} observation marks a hard upper limit on luminosity of ``propeller'' state, however, accretion likely continues even at lower rates as follows from the observed flux variability revealed by \textit{Swift}/XRT at later stages.}
    \label{fig:lc}
\end{figure}

We also monitored the declining phase of the outburst with \textit{Swift}/XRT,
which observed the source at even lower fluxes. The \textit{Swift}/XRT
lightcurve in 0.5-10\,keV band was obtained using the service provided by the
\textit{Swift} data center
\citep{Evans09}\footnote{\url{http://www.swift.ac.uk/user_objects/}} by fitting
spectra of individual observations (see e.g., \citealt{Tsygankov16}). In total
44 observations from MJD~58207 to 58748 with total exposure of $\sim70$\,ks
were considered in this analysis. The absorption column was not well
constrained in all observations, so we fixed it to the interstellar value of
$9\times10^{21}$\,atoms\,cm$^{-2}$ \citep{2013MNRAS.431..394W}. This
approximation does not significantly affect the flux estimates since the final
bolometric flux estimate was obtained by comparing the measured source fluxes
in the 0.5-10\,keV range with the simultaneous \textit{NuSTAR} observations.
This comparison revealed good agreement and that the soft band flux accounts
for $\sim43$\% of the total flux. For observations after MJD~58557.5 (i.e. the
last \emph{NuSTAR} observation) we assumed the continuum to be the same as
revealed by \emph{NuSTAR} and only considered flux as a free parameter. The
resulting lightcurve is presented in Fig.~\ref{fig:lc}. As evident from the
lightcurve, the source brightness continued to decrease after \textit{NuSTAR}
observation, reaching as low as $\sim6\times10^{33}$\lum, however, the counting
statistics in short XRT pointings (1-2\,ks) is not sufficient to detect
pulsations or even robustly detect potential spectral softening. We can not,
therefore, definitively claim that the source continued to accrete also after
\textit{NuSTAR} observation. On the other hand, the observed flux variability
would be hard to explain otherwise, so it is quite possible that accretion
continues even at lower rates than revealed by \emph{NuSTAR}.

Finally, we also obtained the power spectrum of the source to characterize
variability in quiescence. As shown in Fig.\ref{fig:pspec_comparison}, the
power spectrum at low luminosities is consistent with a broken powerlaw with a
break at 0.19(2)\,Hz. The observed aperiodic variability thus also
unambiguously shows that the source continues to accrete. Since the observed
break frequency is higher than the spin frequency by a factor of two, we deduce
that the timescale of the variability relative to the break is not directly
related to the Keplerian timescale at the inner edge of the disk, expected to
be about the spin period at such low luminosity. Unfortunately, the available
statistics does not allow a more detailed timing analysis, e.g., searching for
possible QPOs or investigating the energy dependence of the pulse profiles.

Based on the detection of the pulsations, observed hard energy spectrum, and
aperiodic variability properties, we conclude thus that the source continues to
accrete at very low luminosity.


\subsection{Insight-HXMT}
\begin{figure*}
    \centering
        \includegraphics[width=\textwidth]{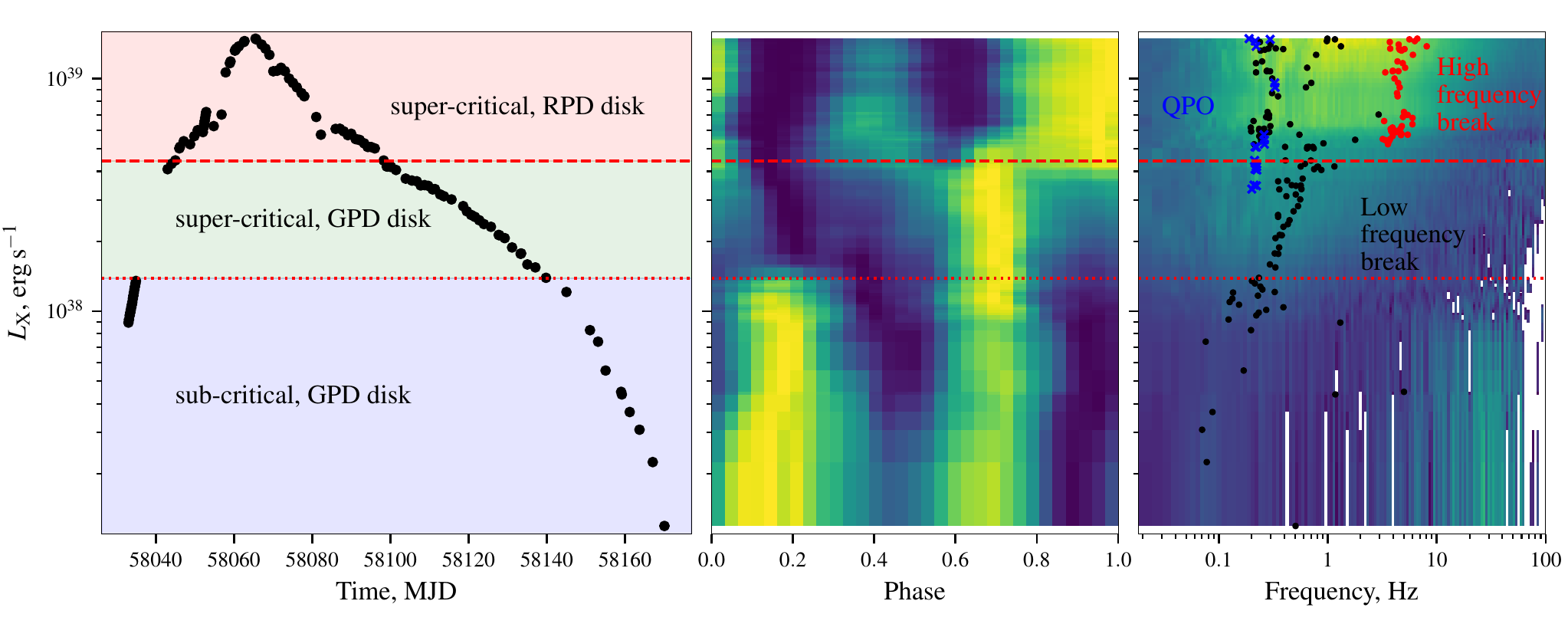}
        
    \caption{Lightcurve of the source as observed by Insight-HXMT (black points) with color-shading showing the different states of the source.
    The relevant transition luminosities marked with horizontal lines in all panels.
    Transition from the sub- to super-critical luminosity is marked by a dramatic change of the pulse profile shape (2nd panel, slices
    at fixed luminosity are pulse profiles
    scaled to the same amplitude to emphasize shape evolution). Transition of the
    accretion disk from GDP to RDP state besides a change in pulse profile shape is also accompanied by a change
    in power spectrum where amplitude of the aperiodic variability increases in 1-10\,Hz frequency range (3rd panel, slices at given
    luminosity are power spectra multiplied by frequency and scaled as square root to emphasize changes in shape).}
    \label{fig:lc_candy}
\end{figure*}

The Hard X-ray Modulation Telescope (HXMT) named ``Insight'' after
launch on June 15 2017 from Jiuquan launch centre, is China’s first X-ray
astronomy satellite \citep{mission_description,hxmt_zhang}. The three main
instruments on-board are the high energy X-ray telescope (HE) boasting
effective area of $\sim$5100\,cm$^2$ between 20-250\,keV, the medium energy
X-ray telescope (ME) operating in 5-30\,keV (effective area 952\,cm$^{2}$), and
the low energy X-ray telescope (LE) operating in 1-15\,keV with effective area
of 384~cm$^2$\citep{hxmt_zhang}.

Insight-HXMT provides an unprecedented view of the phenomenology of the source
at high luminosity. With a total exposure of $\sim835$\,ks accumulated in 98
pointings over the period from MJD~58033 to MJD~58112, observations of \src
constitute the first major observational campaign of the mission. A thorough
analysis of all observations is ongoing and will be presented elsewhere. Here
we focus exclusively on the timing analysis of the HE instrument. With a total
effective area of 5100\,cm$^2$ between 20 and 250\,keV, it is the main
instrument of Insight-HXMT offering high time resolution (0.012\,ms) and low
dead-time even for very bright sources \citep{hxmt_zhang}. This makes the HE an
ideal tool for timing studies, so we used it for timing analysis. We have
verified, however, that similar results can be obtained with ME detector,
although quality of resulting power spectra is lower due to the lower effective
area and shorted good time intervals associated with higher in-orbit
background. For more details on HXMT and performance of individual instruments
please refer to \citep{hxmt_zhang}. The data analysis was performed with {\sc
hxmtdas} v2.01 following the recommended procedures in the user's guide
\footnote{\url{http://www.hxmt.org/images/soft/HXMT\_User\_Manual.pdf}}. More
detail on data analysis for \src can be found in \citep{Zhang19}.

The long-term lightcurve of the source as observed by Insight-HXMT,
\textit{Swift}/BAT and \textit{NuSTAR} is presented in Fig.~\ref{fig:lc}. The
fluxes for Insight-HXMT are estimated based on the broadband spectral analysis
2-150\,keV energy range of individual observations by \cite{Zhang19} and
adopted here from that publication. The values appear to be in line with the
\textit{Swift}/BAT count-rate, which can be robustly converted to luminosity
using the multiplicative factor of $\sim8.2\times10^{38}$ (for a distance of
6.8\,kpc). Note that we omit errorbars for BAT points in Fig.~\ref{fig:lc} for
clarity, but those are rather large, i.e. the light-curve is in excellent
agreement (within uncertainties) with other instruments throughout the
outburst. The \textit{NuSTAR} and \textit{Swift}/XRT fluxes deduced from the
spectral analysis in 3-80\,keV and 0.5-10\,keV energy range are also consistent
with those of Insight-HXMT once the bolometric correction estimated based on
the best-fit model is taken into the account. For instance, for the nearly
simultaneous observation close to the peak of the outburst (i.e. on MJD~58067)
the flux measured by \textit{NuSTAR} and HXMT agree to within $\sim5$\% which
is comparable with flux variations observed within individual observations.

The timing analysis is based on the background-subtracted lightcurves in the
20-80\,keV energy band, with the time resolution of 5\,ms, extracted by
combining detected counts from all non-masked HE instrument modules. Data of
the masked detector are used to estimate the background. For all observation,
the cosmic background component is negligible compared to the source flux and
therefore has been ignored. The lightcurves were corrected for dead-time (not
exceeding 10\%), the effects of the orbital motion of the satellite and for the
binary motion assuming the ephemerides obtained by the Fermi~GBM pulsar team
\footnote{\url{https://gammaray.nsstc.nasa.gov/gbm/science/pulsars/lightcurves/s
wiftj0243.html}}. For each observation, we searched for pulsations, and
determined the period value using the phase-connection technique
\citep{Deeter81}. The spin evolution observed by Insight-HXMT was found to be
consistent with that revealed by the Fermi~GBM \citep{Zhang19}.

To investigate the evolution of the pulse profile shape with luminosity, the
obtained profiles were arranged as a function of the flux \citep{Zhang19} and
aligned with each other using the FFTFIT routine \citep{Taylor92}. Note that
the significant evolution of the pulse profiles with flux implied that we had
to use an iterative procedure, going from low to high fluxes and vice versa,
until the alignment presented in Fig.~\ref{fig:lc_candy} was obtained. We
emphasize the two major changes of the observed pulse profile shape observed at
luminosities of $\sim1.5-4.5\times10^{38}$\lum, in line with findings by
\cite{nicer18}.

\begin{figure}
    \centering
        \includegraphics[width=\columnwidth]{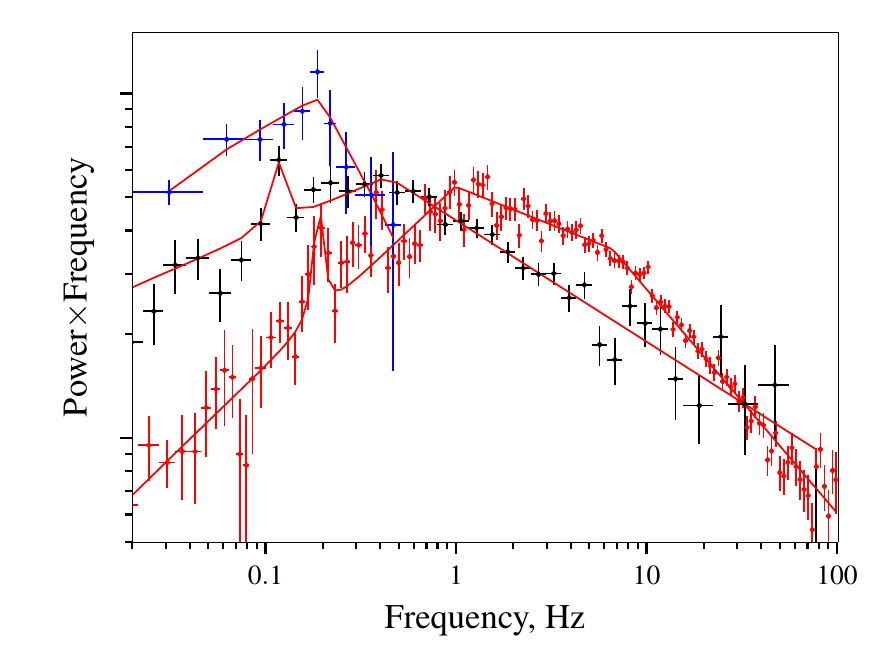}
    \caption{Representative power spectra above (red) and below (black) the
    transition from GPD to RPD state as observed by Insight-HXMT (arbitrary
    scaled and multiplied by the frequency to highlight the changes in shape).
    The low luminosity power spectrum is obtained by combining observations in
    $L_X=2-4\times10^{38}$\lum to improve statistics. The second power spectrum
    is from a single observation slightly above the transition. Estimated white
    noise and pulsed flux had been subtracted in both cases as described in the
    text. Note the QPOs feature around 0.1-0.2\,Hz. Finally, the power spectrum
    of the source in quiescence as observed by \textit{NuSTAR} is also shown
    (blue points). In the latter case pulsations are not subtracted but do not
    contribute significantly to the power spectrum.}
    \label{fig:pspec_comparison}
\end{figure}

High counting statistics also allowed to investigate aperiodic variability in
the source. Here we followed the same approach as \cite{Revnivtsev09} to
suppress the pulsations and enable analysis of the aperiodic noise. In
particular, taking into consideration the average length of good time intervals
and the source spin period, we split lightcurves from each observation in
segments of $\sim300$\,s corresponding to 30 spin cycles. Each segment was then
folded with the spin-period determined for a given observation to obtain an
average pulsed lightcurve, which was then subtracted from the observed
lightcurve. The power spectra of the resulting lightcurves was then obtained
using the \emph{powspec} program by averaging the power spectra of individual
segments. Examples of representative power spectra at different luminosities
are shown in Fig~\ref{fig:pspec_comparison}. The luminosity dependence of the
power spectrum and pulse profiles are better illustrated in
Fig.~\ref{fig:lc_candy} where two distinct regions can be identified.

\begin{figure}
    \centering
        \includegraphics[width=\columnwidth]{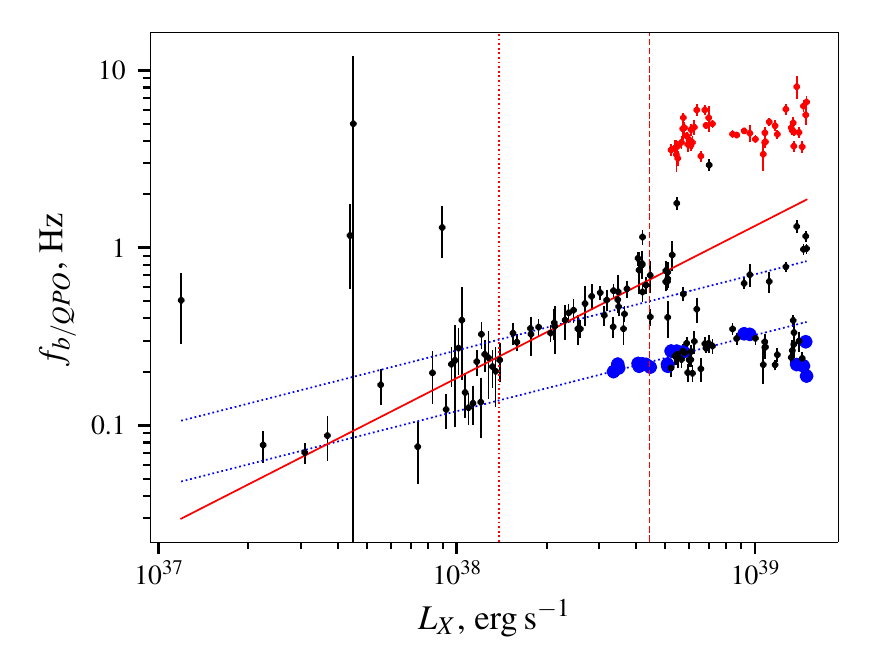}
    \caption{Frequency of the break in the power spectra of the aperiodic variability as function of flux, based on the 20-40\,keV Insight-HXMT HE lightcurves. The black points indicate either the single break (at lower luminosities), or the lower frequency break when the double broken powerlaw model is required to fit the data. The red points indicate the location of the high frequency break where present. The blue points indicate the QPO frequency if detected. 
    To guide the eye we included the red and blue lines to indicate powerlaws with index 6/7 and 3/7 respectively. The vertical lines indicate the fluxes corresponding to dramatic changes of the pulse profile shape (i.e., same fluxes as in Fig.~\ref{fig:lc_candy}).}
    \label{fig:fbreak}
\end{figure}

At low luminosities the power spectrum is well described by a broken power law
as typical for magnetic accretors \citep{Revnivtsev09,Doroshenko14}. Besides
that, some observations reveal low-frequency quasi-periodic oscillations with
frequency of 0.1-0.2\,Hz, as already reported by \cite{nicer18}. The dependence
of the QPO frequency on the flux has not previously reported, likely due to the
shorter duration of NICER observations and more complex pulse profile shape
which complicate subtraction of the pulsations and detection of QPOs. On the
contrary, the power spectra obtained with Insight-HXMT reveal weak QPOs with
luminosity-dependent frequency from $\sim50$\,mHz to $\sim200$\,mHz. We notice,
however, that in several of the observations it's difficult to distinguish the
feature from the remaining variability associated with the imperfect
subtraction of pulsations which makes assessment of the QPO significance rather
complicated. On the other hand, without subtraction of the pulsations QPOs are
not detected at all as the power spectrum is completely dominated by pulsed
flux. We can not exclude, therefore, that apparent presence of QPOs is
associated with imperfect subtraction of the pulsations. The dynamical power
spectrum (see Fig.~\ref{fig:lc_candy}) reveals, however, that the QPO frequency
changes with luminosity, which points to the physical nature of the feature.
Extrapolation of QPO frequency observed by HXMT to lower luminosities
$L_X\sim10^{37}$\lum where similar feature at 50-70\,mHz was reported by
\cite{nicer18} also gives consistent values as can be seen in
Fig.~\ref{fig:fbreak} although no significant QPOs could be detected at this
flux level with HXMT. We note that similar features have been reported for
other X-ray pulsars \citep{Finger96}.

At higher luminosities, i.e., at high accretion rates, the observed power
spectra are different from those at lower luminosities. In particular, a
double, rather than single, broken power law shape is necessary to model the
power spectra spectra as shown in Fig.~\ref{fig:lc_candy},
\ref{fig:pspec_comparison}. We emphasize that this striking change occurs at a
luminosity coinciding with the transition of the observed pulse profile shape,
i.e., at $\sim4\times10^{38}$\lum.

To quantify the evolution of the observed power spectrum, and particularly of
the break frequencies with luminosity, all power spectra were converted to a
format readable by {\sc xspec} (see \cite{Ingram12} for details). The spectra
were then approximated using the double broken powerlaw model implemented as
the \emph{bkn2} model in {\sc xspec}. Subtraction of the pulsations altered the
expected white noise level, which was thus determined by including an
additional powerlaw component with flat power index in the model. For each
observation, we started the modeling by fixing the first break of the
\emph{bkn2} model to the spin frequency, and the second break to 10\,kHz (i.e.,
far above the Nyquist frequency) to mimic the single broken powerlaw typically
observed from other pulsars. We also included a lorentzian line with width
fixed to 0.1 of its central frequency and arbitrary normalization to account
for possible QPOs in all observations. The relative width of the feature was
fixed to a value chosen based on the analysis of several observations where the
width of the QPO feature could be well constrained. The central frequency of
the QPO was then searched between 50 and 200\,mHz, and the feature was finally
included in the fit if the fit statistics improved significantly (i.e. by more
than $2\sigma$ as calculated using the f-test) for any of the trial
frequencies. The same procedure was repeated for the high-frequency break,
which was searched in range between 0.1\,Hz and 1\,kHz. The results are
presented in Fig~\ref{fig:fbreak}.

 We note that the QPO frequency is below the lowest break frequency, by a
factor of $\sim2-3$ similar to what reported for other sources
\citep{Finger96,Revnivtsev09}. The dependence on luminosity $f_{QPO}\propto
L_X^{3/7}$ is also similar, and in fact, consistent with what predicted from
theory if one assumes that the QPO is associated with the Keplerian timescale,
or with the beat frequency between the Keplerian frequency at the inner disk
edge and the neutron star's spin frequency \citep{Finger96}.

We note that in some of the observations it is difficult to constrain the
frequency of the low frequency break due to the presence of the QPOs, and QPO
harmonics at similar frequencies. This is particularly true for brighter
observations where QPOs are more prominent. The scatter of the lower frequency
break fit values at high luminosities shown in Fig~\ref{fig:fbreak} might thus
be simply related to the stability of the fit rather than to some physical
variability. The frequency of the break is also poorly constrained by
Insight-HXMT at low fluxes since the relatively low statistics makes it hard to
distinguish the break from the variable QPO reported by \citep{nicer18} at
$50-70$\,mHz. Finally, it is clear that in the transition region we misidentify
the two breaks for some observations.

Given these limitations, it is hard to draw any robust conclusions regarding
the luminosity dependence of the low frequency break, whether it stays the same
throughout the outburst, or drops to lower frequencies above the transition.
What is clear, however, is that a sharp change in the observed aperiodic
variability properties, coincident with the change of the observed pulse
profile shape, occurs at $L_{\rm X}\sim4-5\times10^{38}$\lum (see
Fig.~\ref{fig:lc_candy}), so we have to conclude that the two transitions are
physically related.

\section{Interpretation and discussion}
\label{sec:disc}
In this section we summarize and interpret the observational results presented
above and in the literature. In particular, we argue that non-detection of the
propeller transition in quiescence implies a relatively low magnetic dipole
field for the pulsar. In this case the accretion disk, at higher accretion
rates, can extend deep into the magnetosphere and has a small inner radius. The
energy release within the disk must in this case be substantial, and in fact,
sufficient for the transition of the inner disk regions to the RPD state. The
transition takes place around MJD~58045 and 58098 in rising and declining parts
of the outburst respectively, and thus is likely responsible for the observed
power and energy spectra, and the pulse profile changes. On the other hand,
changes in source hardness and pulse profile shape reported by
\cite{Tsygankov18} and \cite{nicer18} at slightly lower luminosity (i.e. around
MJD~58035 and 58139 respectively) can in this case be readily associated with
the onset of accretion column. We show that both transitions, the limit on
``propeller'' transitional luminosity, and observed spin-up rate can only be
reconciled if the magnetic field of the neutron star is comparatively weak.
Below we discuss our interpretation in more details.

\subsection{Non-transition to the propeller regime}
As demonstrated above, the source does not enter the ``propeller'' regime even
in quiescence and continues to accrete at fluxes down to at least
$5.5\times10^{-12}$\flux. For the accretion to continue, the source luminosity
must be larger than the propeller luminosity \citep{Tsygankov18}:
\be
L_{\rm prop}\le2\times10^{35}k^{7/2}B_{12}^2\,\mathrm{erg\,s}^{-1}
\label{eq:lprop}
\ee
for standard neutron star parameters. The coupling constant $k$ here accounts
for the effective magnetosphere size compared to the Alfv\`en radius. For the
assumed distance of 6.8\,kpc the accretion flux observed by \emph{NuStar}
implies
$L_{\rm prop}\le3\times10^{34}$\lum, a factor 20
lower than reported by \citep{Tsygankov18}. The lowest luminosity observed by
is another factor of five lower with $L_{\rm prop}\le6\times10^{33}$\lum. Under
the same assumptions (i.e. $k=0.5$), this would imply factor of 4-10 lower
field, i.e. $B_{12}\sim0.6-1.4$, which is rather low compared to other pulsars.
Of course, as already mentioned above, the XRT data do not allow detection of
the pulsations, so that conclusions comes with a caveat. Moreover, as discussed
by \cite{Tsygankov18}, this estimate depends strongly on the rather uncertain
value of the coupling constant, which has thus to be considered as a parameter.
In the context of current work it is, however, more relevant to determine the
condition for the onset of the propeller stage from the comparison of the
co-rotation radius $R_c=(GM/\omega^2)^{1/3}$ with the effective magnetospheric
radius $R_m=kR_A\propto\dot{M}^{2/7}$. This implies $R_m\le
R_c\simeq7.7\times10^8$\,cm in quiescence, so the magnetosphere must be smaller
than $R_m\le7.7\times10^{8}(L_x/L_{\rm prop})^{-2/7} \sim 3-7\times10^7$\,cm at
luminosities 1.5-4.5$\times10^{38}$\lum where transitions in pulse profile
shape and power spectrum take place, and $\sim2-3\times10^7$\,cm close to the
peak of the outburst.

\subsection{Accretion disk at high luminosities}
This conclusion has important implication since local temperature and energy
release rate within the accretion disk increase closer to the compact object.
For highly magnetized neutron stars the magnetosphere normally truncates the
disk far away from the compact object, so energy release within the disk can be
ignored, the gas pressure dominates and the disk remains thin. The situation in
\src is, however, quite different due to the extremely high accretion rate and
small magnetosphere. The boundary between gas pressure and radiation dominated
zones can be estimated from the balance between gas and radiation pressures in
the disk \citep{Juhani19}:
\be
\label{eq:r_ab}
R_\text{AB} = 10^7 m^{1/3} \dot{M}_{17}^{16/21} \alpha^{2/21}\sim4.7\times10^8\ \text{cm},
\ee
close to the peak of the outburst. Here $\alpha \lesssim 1$ is the viscosity
parameter of a Shakura-Sunyaev disk, $m$ is the mass of the neutron star in
units of solar mass, and $\dot{M}_{17}$ is the accretion rate in units of
$10^{17}$\,g\,s$^{-1}$. Given the estimate of the magnetosphere size obtained
above, it is thus clear that substantial part of the disk is likely to be in
the RPD regime at least close to the peak of the outburst.

Indeed, the accretion rate corresponding to RPD transition can be estimated by
equating the transition radius to the magnetosphere size estimate obtained
above, which yields $\dot{M}_{17}\sim14$, i.e. significantly below the peak
accretion rate. We can also explicitly compare it with the standard
magnetosphere radius $R\subt{m}$
\citep{Lamb1973,King2002,Andersson2005,Juhani19}:
\be
\label{eq:r_m}
R\subt{m}= 2.6\powten{8}\ k \ m^{1/7} R_{*,6}^{10/7} B_{12}^{4/7} L_{37}^{-2/7} \text{cm},
\ee
which results in \citep{Juhani19}:
\be
L_\text{AB} = 3\powten{38} k^{21/22} \alpha^{-1/11}  m^{6/11} R_{\text{*,6}}^{7/11} B_{12}^{6/11}\
\text{erg s$^{-1}$.}
\label{eq:lab}
\ee
Dependence of the transitional luminosity on $k,B_{12}$ and $\alpha$ is
comparatively weak, so for any meaningful values the transitional luminosity to
an order of magnitude is $L_{AB}\sim\times10^{38}$\lum, i.e. below the
outbursts peak luminosity.

Moreover, given that the propeller transition was actually not detected, the
estimate of the magnetospheric radius presented above is only an upper limit,
i.e. the accretion disk can, in fact, extend even closer to the neutron star
when radiative pressure becomes dynamically important. The corresponding
characteristic spherization radius is given by \citep{shakura73}:
\be
R_{\rm sp}=\frac{\dot{M}\sigma_T}{4\pi m_pc}\sim10^5\dot{M}_{17}\text{\,cm}
\ee
or $\sim1.9\times10^{7}$\,cm close to the peak of the outburst. While somewhat
smaller than the aforementioned lower limit on the expected magnetosphere size
at outbursts peak, it is clear that as the inner disk radius approaches the
spherization radius, its thickness can not be neglected anymore. Furthermore,
the estimate above neglects additional energy input associated with irradiation
of the disk by compact object and interaction of the accretion flow with the
magnetosphere, so changes in disk structure can be anticipated for larger radii
\citep{Chashkina17}. We conclude, therefore, that non-detection of the
propeller transition in quiescence directly implies that the accretion disk
transitions to RPD state during the outburst, which must have some implications
on structure of the accretion disk and associated observables.

\subsection{Observational evidence for changes in disk structure}
From an observational point of view, the transition to RPD state and
thickening (or eventual spherization) of the accretion disk can be expected to
affect the velocity of matter within the disk and its inner radius
\citep{Chashkina17}. These factors can be expected to affect the geometry of
the accretion flow, and, as a consequence, the observed X-ray spectrum, pulse
profiles, and aperiodic variability properties of the pulsar. Moreover, at this
stage the X-ray emission from the disk itself may become observable. Below we
illustrate that this indeed appears to be the case, and thus argue that
observations confirm presence of a thick RPD disk in \src.

\paragraph*{Spectral transitions, thermal emission from the disk, and onset of the accretion column}
Spectral transitions during the outburst have been discussed by \citet{nicer18} who used NICER and GBM hardness ratios to argue for spectral changes. 
\begin{figure}
    \centering
    \includegraphics{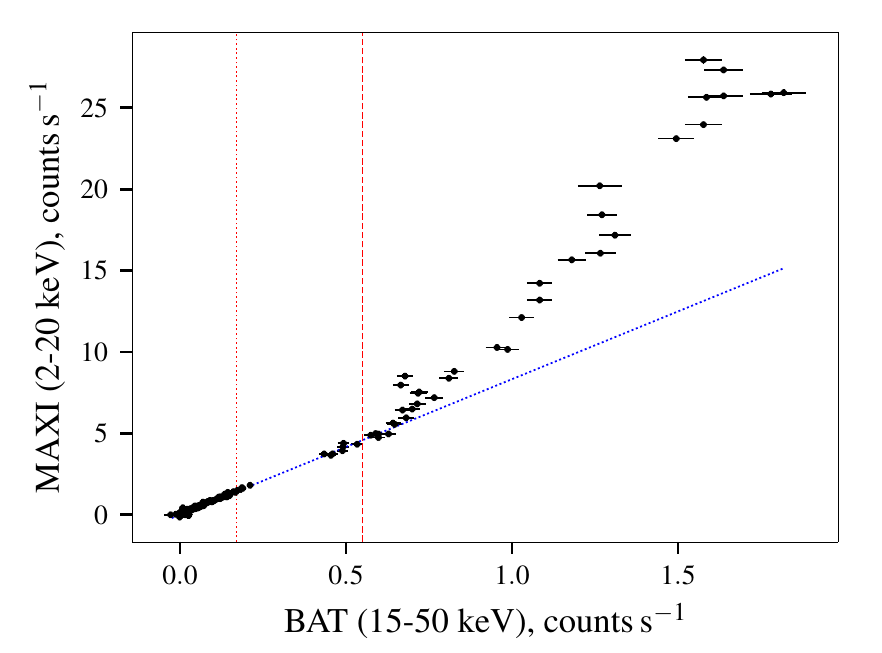}
    \caption{Comparison of count-rates in the soft (2-20\,keV) and hard (15-50\,keV) energy bands based on daily lightcurves by the \textit{Swift}/BAT and MAXI missions. At lower fluxes the two bands are well correlated (blue dotted line), however, this is not the case at higher luminosities, where the brightness in the soft band substantially increases. The vertical lines correspond to the transitional luminosities as observed by Insight-HXMT.}
    \label{fig:bat_maxi}
\end{figure}
In particular, the transition to the super-critical accretion regime has been
suggested to occur at $L_{\rm X}\sim10^{38}$\lum based on the observed
turn-over in the hardness-intensity diagram and pulse profiles in the soft band
\citep[see, i.e. discussion and Fig.~A1]{nicer18}. Same authors also noted
slightly higher transitional luminosity when \textit{Fermi}/GBM colors are
considered, which was attributed this to complex dependence of the source
spectrum and considered both events as single transition from sub- to
super-critical accretion associated with onset of an accretion column.

We note, however, that both transitions turn out to be coincident in luminosity
with the pulse profile changes detected by Insight-HXMT in the hard band at at
luminosities of $\sim1-2$ and $\sim4-5\times10^{38}$\lum. The Insight-HXMT
observes both transitions in a single energy band, which rules out suggestion
by \cite{nicer18} and implies that both events actually took place. This
conclusion is confirmed by the observed spectral evolution of the source
strongly and difference in the observed variability properties described below.

Let us first discuss the spectral evolution. A prominent soft component was
reported by \cite{Tao19} based on the analysis of \textit{NuSTAR} spectra close
to the peak of the outburst. \cite{Tao19} attributed this component to the
emission from the accretion column and an outflow, presumably powered by
accretion from super-Eddington accretion disk. \cite{Tao19} did not estimate
the luminosity corresponding to the appearance of this component, however, it
can be estimated by direct comparison of the soft and hard light-curves as
observed by \textit{Swift}/BAT and MAXI monitors in 15-50\,keV and 2-20\,keV
energy bands. As illustrated in Fig.~\ref{fig:bat_maxi}, the count-rate in the
soft band substantially increases above certain luminosity. Given the evolution
of the spectrum with luminosity reported by \cite{Tao19}, observed spectral
softening is clearly related to the enhancement of the soft component
identified in broadband spectral analysis.

We note that the observed temperature and luminosity (i.e., emission region
size) of the soft component reported by \cite{Tao19} are consistent with
blackbody-like emission of a thick super-Eddington disk truncated at
$\sim10^7$\,cm from the neutron star, in agreement with the estimates of the
inner disk radius discussed above. The appearance of the soft component
coincides with the higher-luminosity transition in pulse profile shape and
aperiodic variability properties revealed by Insight-HXMT, and therefore has to
be related to changes of the accretion disk structure. Given that the only
change expected at this luminosity is the transition of the disk to the RPD
state, we conclude that the observed evolution of spectral and timing
properties $\sim4-5\times10^{38}$\lum is indeed associated with such transition.

The lower luminosity transition can then be readily associated with onset of an
accretion column as suggested by \cite{nicer18,Tsygankov18}, and can be used to
estimate magnetic field of the neutron star. We note that no changes of the
power spectrum associated with this transition are observed, i.e. it is likely
related to the emission region itself rather than the disk. The corresponding
transition luminosity can be estimated $\sim1.5\times10^{38}$\lum based on
evolution of the pulse profiles of the source in hard band revealed by
Insight-HXMT. That is slightly higher, but consistent with
$0.2-1.1\times10^{38}$\lum reported by based on the hardness evolution in the
soft band.

\paragraph*{Features and origin of the observed power spectrum}
The observed flux variability is induced by local accretion rate fluctuations
occurring throughout the accretion disk at timescales related to local
Keplerian timescale \citep{Lyubarskii97} and results in the observed powerlaw
type spectra for flux variability when integrated over the disk. The disruption
of the disk by rotating magnetosphere imposes a break with frequency correlated
with accretion rate and proportional to the Keplerian frequency at the
magnetosphere, which can be used to estimate the inner disk radius
\citep{Revnivtsev09}. At lower luminosities the power spectrum of \src appears
quite similar to that of other X-ray pulsars and magnetic accretors in general
\citep{Revnivtsev09,Suleimanov19}. Indeed, a broken powerlaw power spectrum
with the break correlated with flux, and a QPO with a factor of $\sim2.5-3$
lower frequency are observed in \src similarly to other ``normal'' X-ray
pulsars \citep{Finger96}.
However, above $L_{\rm X}\sim4-5\times10^{38}$\lum, the power spectrum changes
qualitatively, and a second break at higher frequencies appears.

As already mentioned, the transition is also accompanied by a dramatic change
of the observed pulse profile shape around the same luminosity. Given that the
aperiodic variability originates in the accretion disk, it is clear that
simultaneous change of the pulse profile shape and of the power spectrum must
be triggered by a major change in the accretion disk structure rather than that
of emission region, i.e., by the disk transition to the RPD state. Detailed
modeling of the observed evolution of the power spectrum is beyond the scope of
the present work. Below we only discuss it qualitatively in context of the
aperiodic variability properties of RPD disks already discussed in the
literature.

The main open problem to interpret aperiodic variability lies in the not yet
understood origin of the timescales on which accretion rate fluctuations occur
within the disk. \cite{Revnivtsev09} suggested these occur at local Keplerian
timescale. On the other hand, \cite{Mushtukov19} suggested that the dynamo
timescale, expected to be proportional to the local Keplerian timescale,
appears to be a better justified assumption from a physics point of view. In
either case, however, the break frequency in the power spectrum originating in
a truncated disk is expected to correlate with the Keplerian frequency at the
magnetosphere. The exact relation between the two frequencies is still unclear
and subject of theoretical investigations, which hampers quantitative
predictions.

Qualitatively, however, it is reasonable to assume that deviations from the
Keplerian motion within the disk are expected to affect the timescale of
fluctuations, and thus the observed power spectra. Interaction of an RPD disk
with magnetosphere has been considered by \citep{Chashkina17}, who found that
indeed the rotation law in RPD zone differs from the Keplerian by several
percent. More importantly, however, transition to RPD zone alters also the
effective magnetosphere radius, and overall disk structure. That is
particularly relevant if we consider that fluctuations are likely to occur on
dynamo timescale, which also depends on other disk properties such as viscosity
and vertical scale \citep{Mushtukov19}. Both are expected to change upon the
RPD transition, and so it can be expected to alter the emerging power spectrum.

\cite{Juhani19} used this argument to interpret the observed peculiar power
spectrum of ``bursting pulsar'' GRO~J1744$-$28 where a prominent high frequency
noise appears at super-Eddington luminosities. This noise component was
associated with the additional variability induced by the RPD part of the disk,
which was also suggested to suppress the low frequency variability from outer
disk parts. Similar considerations might be qualitatively used also for \src,
where lower noise frequency could be due to the larger magnetosphere
(\gro\xspace has an order of magnitude lower field). However, a detailed
modeling of the power spectrum that takes into account differences in magnetic
field and spin frequency is required to assess whether this scenario is viable.

Alternatively, the appearance of the second break could be related to the
propagation of the emission from the pulsar through the optically thick
envelope expected to enclose a large part of the magnetosphere at high
luminosities \citep{Mushtukov19a}. Finally interaction of the disk with the
quadrupole field component enabled by change of the disk structure could also
explain the emergence of the second break \citep{Juhani19}. The multipole
nature of the magnetic field in the vicinity of the neutron star's surface
could also explain the non-detection of the cyclotron line, which in principle
shall fall within the energy range of \emph{NuSTAR} and \emph{Insight-HXMT}.
Indeed, higher field at the surface might imply a higher line energy, outside
the energy range observed by the HE, or simply could smear a lower energy
feature due significant gradients of the fields across the emission region
expected for quadrupole field configuration. A similar scenario has been
suggested for some ULPs
\citep{2017Sci...355..817I,2017A&A...605A..39T,Middleton19}.
\subsection{Combining all results: the overall scenario}
\begin{figure}
    \includegraphics[width=\columnwidth]{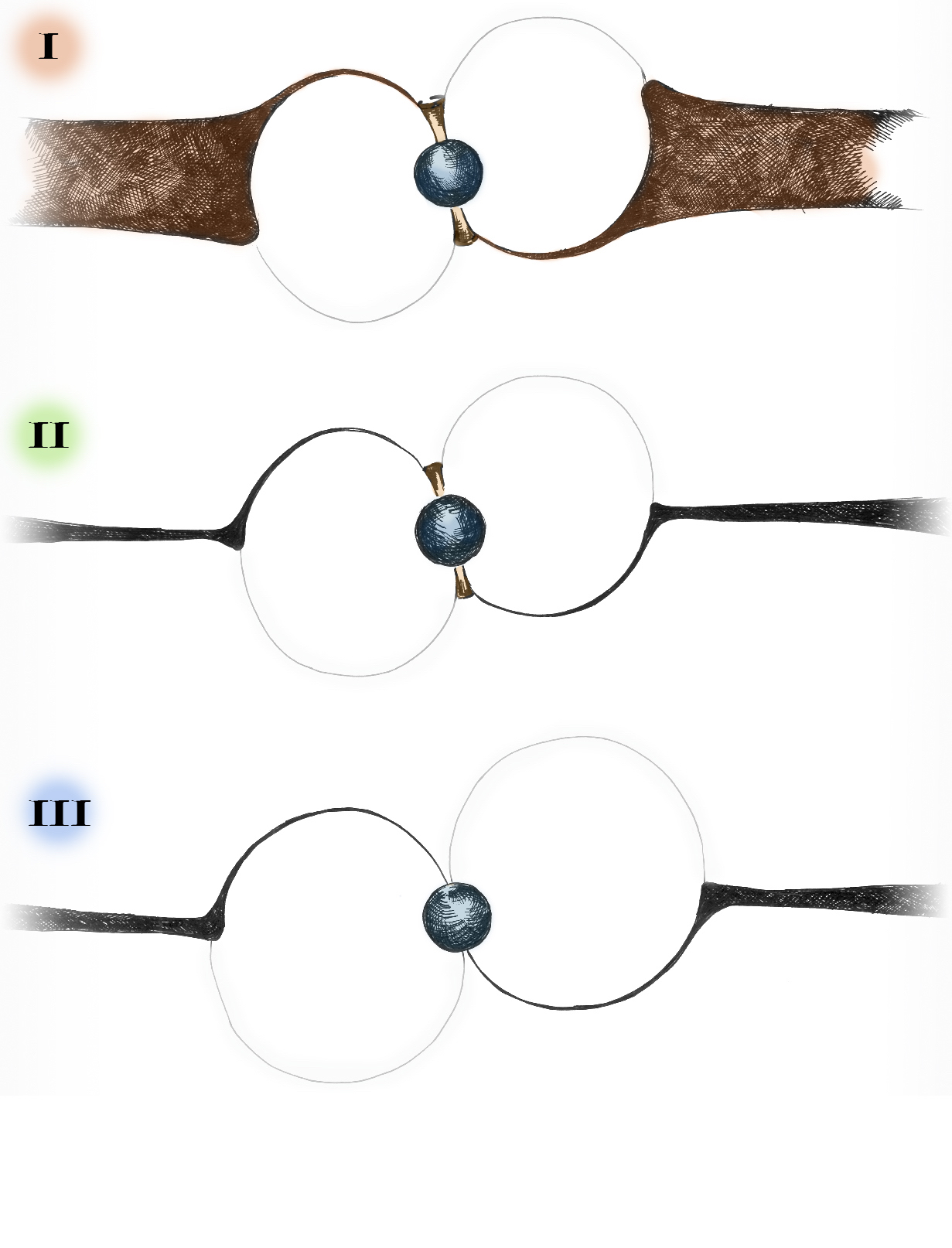}
    \caption{The suggested geometry of the accretion disk and the emission
    region for respective states is also sketched for illustration.}
    \label{fig:cartoon}
\end{figure}

We outline, therefore, the following scenario for the evolution of the
accretion geometry of the source with luminosity. As illustrated in
Fig.~\ref{fig:cartoon}, the source first makes a transition (from III to II)
from the sub- to the super-critical accretion, an accretion column is formed
and the geometry of the emission changes. This change is reflected in the
observed pulse profile shape and luminosity dependence of soft X-ray colours
reported by \cite{nicer18}. In the second, higher luminosity transition the
disk moves from the GPD to the RPD state (from II to I). The disk thickness
and, as a consequence, the geometry of the accretion flow and the emission
region geometry change. This transition is reflected in the change of the power
spectrum, appearance of the strong soft excess in the X-ray spectrum associated
with inner disk regions \citep{Tao19}, and in the detection of radio emission
correlated with X-ray flux attributed by \cite{Eijnden18} to jet formation. The
transition might also be accompanied by a change of the dominant mode of
interaction of the accretion disk with the magnetosphere, i.e., from dipole to
quadrupole.

The main observational arguments in favor of this scenario are the
non-detection of the propeller transition by \emph{NuSTAR} (which implies that
the magnetosphere must be compact both in quiescence and outburst), and the
detection of the transitions in pulse profile and power spectra shape by
HXMT-\emph{Insight} (which can be explained with RPD transition). So far we
tried to keep the discussion independent of the assumed magnetic field of the
source, and have not cross-checked the self consistency of the model beyond
order of magnitude comparisons. With the improved upper limit on the propeller
luminosity it becomes, however, possible to put stronger constrains on the
magnetosphere size both in quiescence and outburst, and thus the magnetic field
of the source.

Indeed, as discussed by \cite{Doroshenko18}, the observed high spin-up rate at
outbursts peak imposes a lower limit on the effective magnetosphere size. On
the other hand, the lack of transition to the propeller state in the
quiescence, imposes an upper limit, and so combining the two turns out quite a
powerful tool to actually measure the size of the magnetosphere. Indeed,
assuming that magnetosphere size in quiescence $R_m$ is close to the corotation
radius $R_c$ and standard scaling for the magnetospheric radius, we can limit
the size of the magnetosphere at any point as
\be
kR_m\le R_c\left(\frac{\dot{M}}{\dot{M}_{\rm prop}}\right)^{-2/7}\sim{7.7\times10^8}\left(\frac{\dot{M}}{\dot{M}_{\rm prop}}\right)^{-2/7}\rm{\,cm,}
\ee
where $\dot{M}_{\rm prop}$ denotes the accretion rate corresponding to the transition to propeller.
At the same time, as discussed by \cite{Doroshenko18}, the magnetosphere must be sufficiently large to explain the observed spin-up rate:
\be
kR_m\ge\left(\frac{I\dot{\omega}}{\dot{M}}\right)^2\frac{1}{GM},
\ee
where $\dot{\omega}$ is the observed spin-up rate at given accretion rate $\dot{M}$. Here we ignore completely any braking torques, so this is an absolute lower limit on the magnetosphere size. Now we can combine the two equations inserting the appropriate numerical values. Considering the uncertainty in the magnetospheric radius dependence on the accretion in the RPD state \citep{Chashkina17}, it makes sense to limit our comparison to the GPD regime where the highest observed spin-up rate is $\dot{\omega}\sim3.3\times10^{-10}$ rad\,s$^{-1}$ at $\dot{M}\sim4.5\times10^{18}$\,g\,s$^{-1}$ \citep{Doroshenko18}. We thus obtain for the same accretion rate
\be
2.82\times10^7{\rm\,cm}\le kR_m \le 4.93\times10^7{\rm\,cm},
\ee
and 
\be
2.82\times10^7{\rm\,cm}\le kR_m \le 3.13\times10^7{\rm\,cm},
\ee
assuming the lowest \textit{NuSTAR} and \textit{Swift} fluxes as limits for the
propeller luminosity. In both cases the limits are consistent with each other,
and the effective magnetosphere size is at least factor of five lower than the
RPD transitional radius as per Eq.\ref{eq:r_ab}. In other words, the observed
spin-up rate is consistent with the suggestion that the disk undergoes the RPD
transition. We note also that the accretion luminosity can not be much lower
than that observed by XRT, as in this case transition to propeller is
inevitable. It can not be excluded, therefore, that variability observed by XRT
actually is associated with unstable accretion around the propeller luminosity.
Nevertheless, assuming $kR_m=3\times10^{7}$, standard definition of
magnetospheric radius, and, as usual, $k=0.5$, this translates to
$B_{12}\sim0.16$, that is an extremely weak field for an X-ray pulsar. This is
highly unlikely, and thus we have to conclude that the accretion disk in fact
penetrates much deeper in the magnetosphere than commonly assumed.

This conclusion can be verified, to some extent, by comparing the observed
luminosity of the second transition, i.e. from sub- to super- critical
accretion regime with theoretical predictions. As already mentioned, the
observed transitional luminosity is $L_{\rm crit}\sim1.5\times10^{38}$\lum,
which can be compared with the same relation by \cite{Becker12} as
\cite{nicer18}. Assuming $kR_m\sim3\times10^7$\,cm we arrive thereby to the
same value of $B_{12}\sim1.4\times10^{-35}L_{\rm crit}^{-15/16}\sim8.7$ as
\cite{nicer18}. In this case $k\sim0.1$ is required to explain the estimated
effective magnetosphere size. On the other hand, theoretical predictions of
critical luminosity are rather uncertain on their own, since the critical
luminosity is affected by assumed emission region geometry, particularly area
of the hotspots on the surface of neutron star. This area is defined by the
geometry of the accretion flow, which is believed to be defined by the
effective magnetosphere size \citep{Mushtukov19}. This dependency is ignored by
\citet{Becker12}, and here estimate by \cite{Mushtukov15} is more appropriate.
Numerical evaluation using this model gives slightly lower field of
$B_{12}\sim3-5$ and $k\sim0.1-0.2$ depending on whether the accretion column is
assumed to be filled or hollow. We emphasize that despite this uncertainty,
both models predict rather moderate magnetic field, and allow to explain
self-consistently both transitions observed by Insight-HXMT, non-detection of
the propeller transition, and the observed spin-up rate. However, small values
of coupling constant $k$ are clearly preferred.

The RPD transitional luminosity predicted for the estimated values of $k$ and
$B_{12}$ appears by factor of two lower than observed, and in fact, more
consistent with the observed luminosity of first transition
$L_{AB}\sim2\times10^{38}$\lum, i.e. the transition to RDP regime occurs
simultaneously with onset of an accretion column, which is a surprising
coincidence. In principle, the effects of RPD transition could be expected to
become apparent at higher luminosities when substantial part of the disk
transitions to RPD state. Furthermore, we only considered a fixed assumed
distance, which is also rather uncertain and decreasing the distance to lower
limit of $\sim5$\,kpc would improve the agreement. Considering also that all
estimates above are actually rather rough, the factor of two agreement can,
nevertheless, be considered excellent. The important point also is that the
scenario remains self-consistent also considering other constrains on the
magnetic field (i.e. spin-up rate and transition luminosity from sub- to
super-critical accretion), and the RDP transition is still expected to take
place at around the observed luminosity.

\section{Summary and conclusions}
\label{sec:sum}
The surprise outburst of \src allowed us for the first time to study accretion
physics at ULX-like accretion rates, and in particular to discover the
transition of inner disk regions to the RPD state. As discussed above,
uncertainty in distance and magnetic field of the neutron star complicates
interpretation of observational findings, nevertheless, we are able to conclude
that in this scenario the magnetic field must be comparatively weak to
coherently explain all of the observed phenomenology. In particular, the
observed continued accretion at lowest luminosity
$L_x\sim0.6-3\times10^{34}$\lum, transition from the sub- to the super-
critical regime and onset of the accretion column at
$L_x\sim1-2\times10^{38}$\lum, transition to the RPD state at the highest
luminosity $L_x\sim4.4\times10^{38}$\lum accompanied by the appearance of the
strong soft excess in X-ray spectrum, and the observed spin-up rate of the
neutron star throughout the outburst. Considering all observables, we conclude
that the dipole component of the magnetic field in \src must in this case be
$B\sim3-9\times10^{12}$\,G, and likely at the lower limit of this range, i.e.,
in range typical for accreting pulsars. Even so, the source has been able to
reach ULP luminosity levels while still pulsating. This conclusion is in line
with recent field estimates
\citep{2015RAA....15..517T,2017MNRAS.465L...6C,2017ApJ...838...98X,2019MNRAS.482
.4956T} for extra-galactic ULPs, where magnetar-like like fields were initially
suggested \citep{2015MNRAS.454.2539M}.

We note that our conclusions are confirmed by findings of \cite{nicer_iron} who
investigated shape of the iron line and thermal emission from the source based
on \textit{NICER} observations. The iron line was found to broaden with
luminosity suggesting high velocities in inner disc regions and inner disc
radius as small as $\sim5\times10^{7}$\,cm, i.e. consistent with our findings.
Thermal emission from the disc similar to that reported by \cite{Tao19} was
also detected, although interpretation of the broadband continuum, as discussed
by \cite{nicer_iron}, is slightly different in the two cases.
 
It can be possible due to strong multiple components of the magnetic field (see
e.g. \citealt{2017Sci...355..817I}). Another possibility is related to the
geometrical thickness of accretion column: geometrically thin accretion columns
can support larger accretion luminosity (see approximate equation 10 in
\citealt{2015MNRAS.454.2539M}). It is assumed that the geometrical thickness of
a column is determined by the penetration depth of accretion disk into the
magnetosphere. In the paper by \citealt{2015MNRAS.454.2539M} the penetration
depth was taken to be about a geometrical thickness of a disk at the
magnetospheric radius. Because the accretion disk tends to be radiation
pressure dominated and geometrically thick at large mass accretion rates, the
geometrical thickness of accretion column was taken to be large. However, if
the penetration depth of the accretion disk into the magnetosphere is
significantly smaller than its geometrical thickness (it might be due to a
strong radiation force at the inner disk edge), the geometrical thickness of
accretion columns in the paper by \citealt{2015MNRAS.454.2539M} was
overestimated, and, therefore, the maximal luminosity of the columns was
underestimated. In this case the enormous accretion luminosity of ULX pulsars
can be explained without the hypothesis of magnetar-like magnetic fields.

We note also that high magnetic fields inferred for ULPs, and in particular
M82~X-2 \citep{Bachetti,Tsygankov16} largely stem from the assumption that the
disk is truncated at approximately half of the Alfv\'enic radius
\citep{Tsygankov16} and may be overestimated if it is not the case. As
discussed above, the effective magnetosphere size in \src must definitively be
small (with $k\sim0.1-0.2$), and the same possibility has, in fact, been
already theoretically discussed for ULPs \citep{Chashkina17,Mushtukov19a}. This
might be associated with the transition to RPD state, although potential
presence of multipole field components in another luminous X-ray pulsar,
SMC~X$-$3, has been also suggested as a possibility to resolve discrepancy
between various estimates of the magnetic field in ULPs
\citep{2017A&A...605A..39T}. Observations of \src thus might provide the first,
and possibly the only direct confirmation of this hypothesis.
 
Given the extra-galactic origin of ULPs and their low observed fluxes which
complicate detailed observational investigation of these objects, we finally
conclude that \src and similar Galactic sources provide a unique close-up view
on accretion physics in ULPs and represent an ideal playground for testing the
theoretical predictions for accretion physics on magnetized neutron stars.
Observed properties of \src echo the theoretical considerations already invoked
for ULPs, and the detailed analysis of the vast amount of high quality
Insight-HXMT, \textit{NuSTAR}, \textit{NICER}, and VLA observations of this
source will likely be key for tackling the problem of ULPs.

\section*{Acknowledgements}
This work made use of the data from the Insight-HXMT mission, a project funded by China National Space Administration (CNSA) and the Chinese Academy of Sciences (CAS). The Insight-HXMT team gratefully acknowledges the support from the National Program on Key Research and Development Project (Grant No. 2016YFA0400800) from the Minister of Science and Technology of China (MOST) and the Strategic Priority Research Program of the Chinese Academy of Sciences (Grant No. XDB23040400). The authors thank supports from the National Natural Science Foundation of China under Grants No. 11503027, 11673023, 11733009,
U1838201 and U1838202.
This work was supported by the Russian Science Foundation grant 19-12-00423 (VD, ST, AM).
This research has made use of MAXI data provided by RIKEN, JAXA and the MAXI team. We acknowledge the use of public data and products from the \textit{Swift} data archive.
We also thank the {\it NuSTAR} team for approving the DDT observation of \src.
\bibliographystyle{mnras}
\bibliography{biblio}   
\end{document}